\documentclass[ALICE,manyauthors]{cernphprep}
\usepackage[comma,square,numbers,sort&compress]{natbib}
\usepackage{hyperref}
\usepackage{lineno}
\usepackage[all]{nowidow} 
\usepackage{isomath} 
\usepackage{siunitx} 

\usepackage{bm} 
\usepackage{hyphenat} 
\usepackage{xcolor}
\usepackage[T1]{fontenc}
\usepackage{orcidlink}

\newcommand{\tn}[1]{\textnormal{#1}} 
\DeclareMathAlphabet{\mathup}{OT1}{\familydefault}{m}{n} 

\newcommand{\urlss}[1]{{\scriptsize \url{#1}}} 

\newcommand{\Dt}{\mathup{\Delta}} 

\DeclareSIUnit\clight{\text{\ensuremath{c}}} 
\DeclareSIUnit[number-unit-product = ]\percent{\char`\%} 
\newcommand{\tev}{\si{\tera\electronvolt}} 
\newcommand{\gevc}{\si[per-mode=symbol]{\giga\electronvolt\per\clight}} 
\newcommand{\gevcc}{\si[per-mode=symbol]{\giga\electronvolt\per\clight\squared}} 

\newcommand{\particle}[1]{\textup{#1}} 

\newcommand{\proton}{\particle{p}}

\newcommand{\pion}{\particlegr{p}}
\newcommand{\kaon}{\particle{K}}

\renewcommand{\pion}{\ensuremath{\uppi}}

\newcommand{\pionplus}{\ensuremath{\pion^{+}}}

\newcommand{\kaonminus}{\ensuremath{\kaon^{-}}}

\newcommand{\dzero}{\ensuremath{\particle{D}^{0}}}

\newcommand{\pp}{\mbox{\proton\proton}} 

\newcommand{\pt}{p_{\tn{T}}} 
\newcommand{\pts}[1]{p_{\tn{T, }{#1}}} 
\newcommand{\ptdzero}{\pt^{\dzero}} 
\newcommand{\jetch}{\tn{jet ch}} 
\newcommand{\ptjet}{\pt^{\jetch}}

\newcommand{\zg}{z_{\tn{g}}}
\newcommand{\rg}{R_{\tn{g}}}

\newcommand{\nsd}{n_{\tn{SD}}}

\newcommand{\zcut}{z_{\tn{cut}}}

\newcommand{\kt}{\ensuremath{k_{\tn{T}}}} 
\newcommand{\akt}{anti-{\kt}} 

\begin{document}

\begin{titlepage}
\PHyear{2022}       
\PHnumber{171}      
\PHdate{08 August}  

\title{Measurements of groomed\hyp{}jet substructure of charm jets tagged by \dzero{} mesons in proton--proton collisions at $\sqrt{s} = \SI{13}{\tev}$}
\ShortTitle{Measurements of groomed charm\hyp{}jet substructure}  

\Collaboration{ALICE Collaboration\thanks{See Appendix~\ref{app:collab} for the list of collaboration members}}
\ShortAuthor{ALICE Collaboration} 

\begin{abstract}
Understanding the role of parton mass and Casimir colour factors in the quantum chromodynamics parton shower represents an important step in characterising the emission properties of heavy quarks. Recent experimental advances in jet substructure techniques have provided the opportunity to isolate and characterise gluon emissions from heavy quarks. In this work, the first direct experimental constraint on the charm-quark splitting function is presented, obtained via the measurement of the groomed shared momentum fraction of the first splitting in charm jets, tagged by a reconstructed \dzero{} meson. The measurement is made in proton--proton collisions at $\sqrt{s} = \SI{13}{\tev}$, in the low jet transverse-momentum interval of $15 \leq \ptjet < \SI{30}{\gevc}$ where the emission properties are sensitive to parton mass effects. In addition, the opening angle of the first perturbative emission of the charm quark, as well as the number of perturbative emissions it undergoes, is reported. Comparisons to measurements of an inclusive-jet sample show a steeper splitting function for charm quarks compared with gluons and light quarks. Charm quarks also undergo fewer perturbative emissions in the parton shower, with a reduced probability of large-angle emissions.

\end{abstract}
\end{titlepage}

\setcounter{page}{2} 


In hadronic collisions, quarks and gluons (partons) with high transverse momentum ($\pt$) and/or large masses are produced in scattering processes involving large momentum transfers. These energetic partons lose their large initial virtuality by emitting partons in a~cascade process known as a~parton shower, until they eventually form hadrons. Theoretical descriptions of parton showers rely on determining the emission probability of partons, which are described by splitting functions~\cite{Angular_Ordering,ALTARELLIPARISI}. Splitting functions are universal properties of quantum chromodynamics (QCD) that describe how the energy of a parton is shared as it fragments into further partons. The splitting functions of light-flavour (up, down, and strange) quarks and gluons differ due to their different Casimir colour factors~\cite{ALTARELLIPARISI}. 
For heavy-flavour (charm and beauty with masses of \SI{1.27(02)}{\gevcc} and $4.18^{+0.03}_{-0.02}$~\si{\gevcc}, respectively~\cite{PDG}) quarks the splitting functions are strongly influenced by the large quark mass~\cite{HQNLO}. As a consequence, the radiation pattern of the parton shower depends on the type of initiating parton. 

Experimental access to the parton shower can be gained via measurements involving jets, which are collimated bunches of hadrons resulting from the fragmentation of partons scattered in the initial stages of collisions.
Jets represent a~powerful tool for testing perturbative QCD, as their inner structure (substructure) reflects the properties of the underlying parton shower. Measurements of jet substructure have been performed in inclusive jets (with no flavour tagging) to constrain the splitting functions of light quarks and gluons~\cite{CMS2018zgppPbPb,ALICE_SoftDrop,ATLAS_SoftDrop,STAR_SoftDrop}. However, an experimentally clean separation of gluon-initiated and quark-initiated jets has remained challenging, with measurements constraining the splitting functions comprising an admixture of the two subsets.

Recent experimental advances in techniques pertaining to the substructure of heavy-flavour jets~\cite{Deadcone_Leticia,HFsplitting2017} have allowed for the isolation and study of gluon emissions from heavy quarks. This technique has been used to perform a first direct measurement~\cite{alicecollaboration2021direct} of the dead-cone effect~\cite{deadcone} in charm-quark jets. The dead cone corresponds to an angular region along the flight direction of the emitter, with an opening proportional to the ratio of mass to energy of the emitter, within which the probability of emissions is suppressed. The significant size of this region for low-energy heavy-flavour quarks gives rise to mass-dependent effects in the parton shower. In addition to the sensitivity to these mass effects in QCD, heavy-flavour jets represent an experimentally enhanced quark-initiated jet sample, which can be further used to constrain the flavour-dependent properties of QCD emissions arising from different Casimir colour factors.

In this work, the first measurement directly constraining the charm-quark splitting function is reported. This requires access to perturbative splittings (the splitting scale is much larger than the QCD scale) of the charm quark, where the shared momentum fraction of the splittings maps onto the analytical splitting function~\cite{cms-splitting-function,sudakov-safety,SplittingFunctionAndZgLL}. Additionally, measurements of the opening angle of these emissions, as well as the number of perturbative emissions in charm jets, are reported. Measurements of these two observables have also been reported for inclusive jets~\cite{ALICESoftDropNotUnfolded,ALICE_SoftDrop,ATLAS_SoftDrop,STAR_SoftDrop,alice-softdrop-inclusive}, with the opening angle calculated at next\hyp{}to\hyp{}leading logarithmic accuracy~\cite{RgNLL}. The measurements reported in this work are made in the low jet transverse-momentum interval of $15 \leq \ptjet < \SI{30}{\gevc}$, where mass effects are expected to play a~role in the fragmentation of the charm quark and the inclusive sample is mainly comprised of gluon-initiated jets.

Once jets have been identified from the sample of charged particles (reconstructed as tracks) in \pp{} collisions, using the \akt{} algorithm~\cite{antikt}, track\hyp{}based jet\hyp{}substructure observables can be constructed from the jet constituents. To reconstruct the chain of emissions (splitting tree) inside a jet, the jet constituents can be reclustered with the Cambridge–Aachen (C/A) algorithm~\cite{ca}, which is well suited to jet substructure studies because it follows the angular ordering of emissions in QCD~\cite{Angular_Ordering} by clustering the jet constituents based on their angular distance. By following a branch along the splitting tree returned by the reclustering procedure, a sequence of emissions can be studied~\cite{lund-plane}.

Charm jets are identified by the presence of a reconstructed~\dzero{} meson amongst their constituents. The full reconstruction of the $\dzero \to \kaonminus \pionplus$ decay allows for the replacement of the \dzero{} decay products with the four\hyp{}momentum of the \dzero{} meson, prior to jet clustering.
This ensures that the full \dzero{}-meson momentum is contained within the jet cone. As the charm flavour is conserved throughout the jet evolution, the charm quark can be traced as it dynamically evolves in the showering process, by following the branch containing the \dzero{} meson through the splitting tree~\cite{Deadcone_Leticia}.

To reduce the contribution of non-perturbative effects and increase sensitivity to perturbative emissions~\cite{ALICE:Angularities}, the Soft Drop grooming procedure~\cite{SoftDrop_Groomer} is applied, which only selects particular splittings of interest along the followed branch. To do this, each splitting along this branch is tested against the Soft Drop condition, as given by
\begin{equation}
    z \equiv \frac{\min (\pts{1}, \pts{2} )}{\pts{1} + \pts{2}} > \zcut \left(\frac{\Dt R_{1, 2}}{R}\right)^{\beta},
    \label{Eq:Soft_Drop}
\end{equation}
where $\pts{1}$ and $\pts{2}$ are the transverse momenta of the leading and subleading prongs of the splitting, respectively, and $R$ is the jet resolution parameter.
The grooming behaviour is determined by the parameters $\zcut$ and $\beta$, which control the interplay between the shared momentum fraction, $z$, and the aperture angle between the prongs, $\Dt R_{1,2} \equiv \sqrt{(y_{1}-y_{2})^{2}+(\varphi_{1}-\varphi_{2})^{2}}$, where $y$ and $\varphi$ are the rapidity and azimuth of prongs 1 and 2, respectively.
In this work, values of $\zcut = 0.1 $ and $\beta = 0$ are chosen, such that the Soft Drop condition is satisfied if the subleading prong carries at least \SI{10}{\percent} of the sum of the transverse momenta of the prongs, enriching the selection with perturbative splittings. 

Two groomed substructure observables are constructed against the first splitting that satisfies the Soft Drop condition:
$\zg = z$ and $\rg = \Dt R_{1,2}$, as given in Eq.~\ref{Eq:Soft_Drop}. For the given choices of grooming parameters, the shared momentum fraction, $\zg$, converges to the QCD splitting function at sufficiently high jet energies~\cite{cms-splitting-function}. As the emissions are angular ordered, the groomed emission angle $\rg$ characterises the widest emission in the splitting tree which passes the Soft Drop condition and sets the geometrical scale of the jet. This observable is expected to be sensitive to the dead cone of the charm quark at small angles, whilst emissions from gluons, which have a larger Casimir colour factor, are expected to dominate the large-angle region~\cite{OpalQuarkGluon, CMSQuarkGluon}. The number of emissions of the charm quark satisfying the Soft Drop condition, $\nsd$, is also measured by evaluating all splittings along the branch containing the \dzero{} meson. This strongly correlates to the number of perturbative emissions of the charm quark.

The data were collected using the ALICE apparatus at the LHC, during the years 2016, 2017, and 2018. Information about the detector configuration and performance can be found in Refs.~\citenum{ALICE_1,ALICE_2}. The main detector systems used for this work were the central\hyp{}barrel detectors, located at $|\eta| <0.9$ within a~solenoidal magnet and used for charged\hyp{}particle tracking and identification.
These include the Inner Tracking System, the Time Projection Chamber, and the Time\hyp{}Of\hyp{}Flight detector.
The V0 detector, consisting of two scintillator arrays located at pseudorapidities of $2.8 < \eta < 5.1$ and $-3.7 < \eta < -1.7$, was used for the trigger and event selections.
Minimum\hyp{}bias events were selected by requiring a~signal above a~given threshold in both V0 counters.
The analysed data sample consisted of about \num{1.7e9} \pp{} collisions at a centre-of-mass energy of $\sqrt{s} = \SI{13}{\tev}$, corresponding to an integrated luminosity of $\mathcal{L}_{\text{int}}=\SI{29}{nb^{-1}}$.

The \dzero{} mesons were reconstructed in the \dzero{} $\pt$ interval of $5 \leq \ptdzero < \SI{30}{\gevc}$, through the $\dzero\to\kaonminus\pionplus$ (and charge conjugate) decay channel with a branching ratio of \SI{3.95(3)}{\percent}~\cite{PDG}. Background candidates, built from pairs of tracks not corresponding to a~\dzero{} decay, were suppressed by applying geometrical selections on the displaced\hyp{}decay\hyp{}vertex topology, as well as particle identification on the decay-particle tracks~\cite{D0_Old_Paper}. The \dzero{} decay daughters were replaced in the event by the \dzero{}\hyp{}meson four-momentum, which was obtained by summing the four-momenta of these decay daughters. Jet finding was then performed using the FastJet package~\cite{fastjet}, with the \akt{} algorithm, a~resolution parameter of $R=0.4$ and the $E$\hyp{}scheme recombination. The inputs to the jet finder consisted of tracks of charged particles with $\pt \geq \SI{0.15}{\gevc}$ and the reconstructed \dzero{}\hyp{}meson candidate. The full set of quality selections applied to the tracks, including the \dzero{}-meson candidate decay tracks, is given in Ref.~\citenum{D0fragmentation}. The jets were required to be fully contained within the central barrel acceptance, by imposing a pseudorapidity selection of $|\eta| < 0.5$ on the jet axis.
For every \dzero{} meson candidate in an event, jet finding was performed independently, with only the two tracks forming that particular candidate replaced with the sum of their four-momenta prior to the jet finding pass.
Once jet finding had been performed, the \dzero{}\hyp{}tagged jet constituents were reclustered using the C/A algorithm and the $\zg$, $\rg$, and $\nsd$ observables were calculated for each jet. The contribution to the three observable distributions from jets tagged by background \dzero{} candidates that survived the selection was removed via a~sideband subtraction procedure, as detailed in Refs.~\citenum{D0fragmentation,d0-substructure-note}. This procedure provides the true \dzero{}-tagged jet observable distributions in intervals of $\ptdzero$.

Each of these background-subtracted distributions was scaled by the selection and reconstruction efficiency of prompt (charm\hyp{}showering) \dzero\hyp{}tagged jets, estimated in $\ptdzero$ intervals, from Monte Carlo (MC) simulations using PYTHIA~8.243~\cite{MONASH,PYTHIA6,PYTHIA8-2} and GEANT~3~\cite{GEANT3}.
The efficiency\hyp{}corrected distributions were summed over the full $\ptdzero$ range considered for the analysis. In order to study emissions in the charm-quark shower, the contribution from non-prompt \dzero{} mesons (feed\hyp{}down contribution), originating from the decay of beauty hadrons, was estimated using POWHEG~\cite{POWHEG} simulations with PYTHIA~6.425~\cite{PYTHIA6} showering and decays implemented through the EvtGen package~\cite{evtgen}. It was then subtracted from the measured distributions. The magnitude of this non-prompt contribution ranged from \SI{20}{\percent} to \SI{40}{\percent} in most of the $\zg$, $\rg$, and $\nsd$ intervals considered in the analysis. 
The $\ptjet$ distributions and each prompt \dzero\hyp{}tagged jet substructure observable were simultaneously corrected for detector effects via a~two\hyp{}dimensional unfolding, using the iterative Bayesian unfolding algorithm with four iterations~\cite{unfolding}.

The $\zg$, $\rg$, and $\nsd$ jet observables were also measured for inclusive jets. To allow for direct comparisons with the heavy-flavour results, the constraint on the $Q^{2}$ value of the parton scattering implied by the selection of \dzero{} mesons with $\ptdzero \geq \SI{5}{\gevc}$ was mimicked for inclusive jets, by requiring a~leading track with $\pt \geq \SI{5.33}{\gevc}$. This corresponds to the $\pt$ of a~charged pion with the same transverse mass as a~\dzero{} meson with $\ptdzero = \SI{5}{\gevc}$.
In addition, during the grooming procedure, the hardest branch of the splitting tree was followed. In the heavy-flavour sample, following the hardest branch was found to coincide with following the branch containing the \dzero{}, in over \SI{99}{\percent} of cases.
The inclusive\hyp{}jet distributions were then corrected for detector effects via the same unfolding procedure as in the \dzero{}\hyp{}tagged jet case.

Systematic uncertainties related to the reconstruction and identification of prompt \dzero-tagged jets were studied. The stability of the selections applied to the \dzero-meson candidates were estimated by loosening and tightening the topological and particle identification criteria used to identify the \dzero{} decays. The uncertainty pertaining to the signal extraction procedure was estimated by changing the fitting parameters on the invariant\hyp{}mass fits, as well as the widths of the signal and sideband regions used for the sideband subtraction procedure. The uncertainty on the \dzero{} feed-down estimation was obtained by varying the renormalisation and factorisation scales, as well as the mass of the beauty quark, in the POWHEG simulations~\cite{powheg-variations}. The uncertainty on the jet energy resolution was estimated by artificially reducing the tracking efficiency used to generate the unfolding response matrix by $4\%$, in line with the tracking efficiency uncertainty of the ALICE detector. Finally, the uncertainty associated with the unfolding procedure was obtained by varying the number of iterations, the choice of prior, and the ranges of the response matrix. Additionally, variations of the fragmentation properties of the MC simulation used to construct the response matrix were also included. These were obtained by reweighting the response matrix based on the value of the jet-angularity observable~\cite{Angularities} ($\kappa =1$ and $\beta =1$) calculated for each entry, as this observable is sensitive to the transverse momentum and angular distribution of all hadrons in the jet. The weights were devised such that the reweighting procedure transformed the angularity distribution at generator level of the \dzero\hyp{}tagged jets in the response matrix, to match that of a simulated inclusive\hyp{}jet distribution.

The dominant contributions in most of the intervals resulted from variations of the topological selections, signal extraction, and feed\hyp{}down subtraction. For the feed\hyp{}down subtraction uncertainty, the maximum negative and the maximum positive deviations were taken, respectively. For each of the other variation categories, the root mean square of deviations from the central values was calculated and assigned as a~systematic uncertainty, which was then symmetrised around the central value.
The uncertainties from all categories were then added in quadrature to obtain the full systematic uncertainty. The total systematic uncertainties range from \SI{7}{\percent} to \SI{77}{\percent}, from \SI{9}{\percent} to \SI{12}{\percent}, and from \SI{5}{\percent} to \SI{59}{\percent} for the \dzero-tagged jet $\zg$, $\rg$, and $\nsd$ distributions, respectively.

The systematic uncertainties for the inclusive\hyp{}jet distributions were estimated by performing variations of the tracking efficiency and unfolding, in the same way as in the \dzero-tagged jet case. In addition, the $\pt$ selection on the leading track of the jet was also varied in line with the track $\pt$ resolution of the \dzero-decay tracks. The total systematic uncertainties range from \SI{1.4}{\percent} to \SI{4.3}{\percent}, from \SI{2.2}{\percent} to \SI{5.7}{\percent}, and from \SI{3.3}{\percent} to \SI{8.1}{\percent} for the inclusive\hyp{}jet $\zg$, $\rg$, and $\nsd$ distributions, respectively.

\begin{figure}[hpt]
    \centering
    \includegraphics[width=0.49\textwidth]{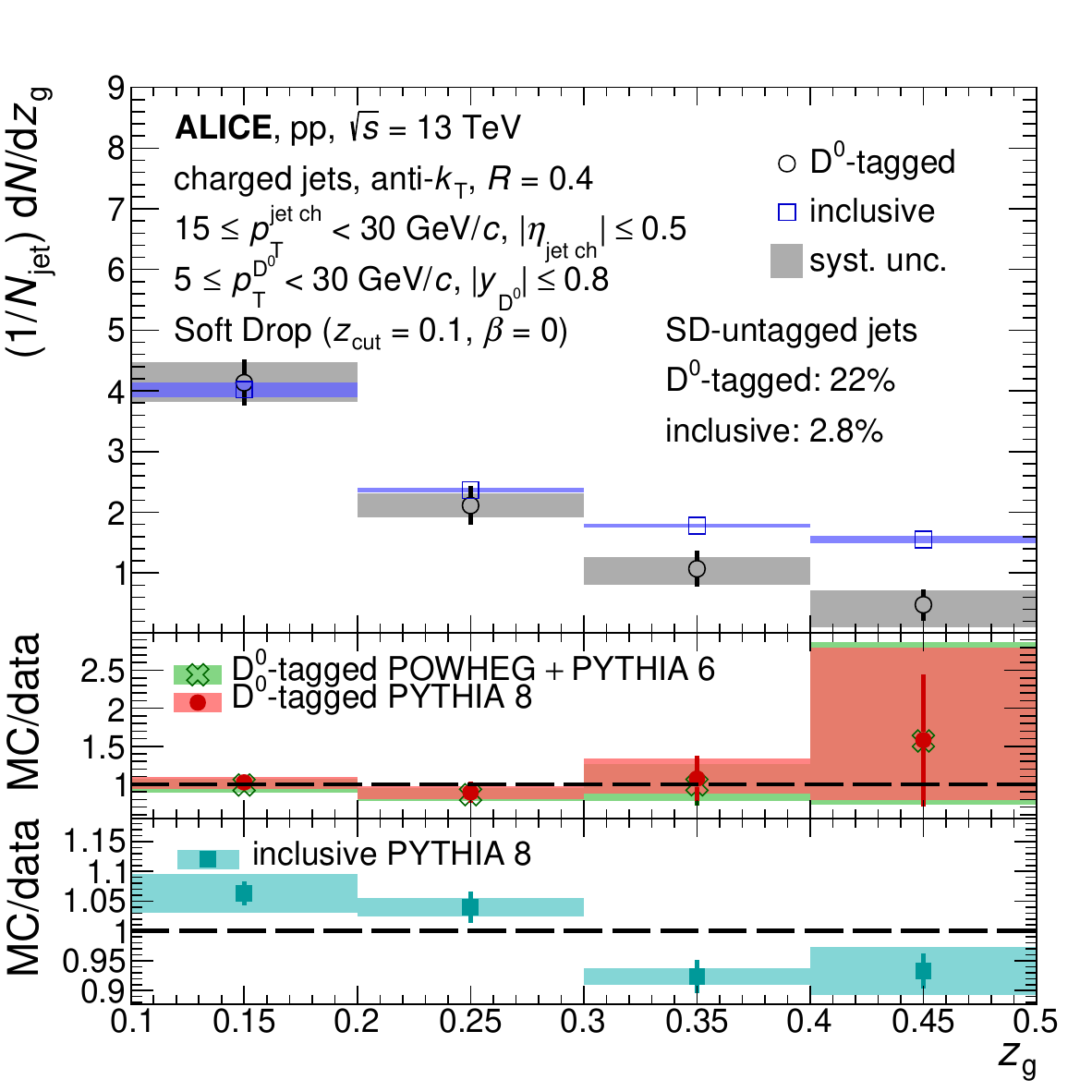}
    \caption{The $\zg$ distribution of prompt \dzero\hyp{}tagged jets compared to that of inclusive jets for $15 \leq \ptjet < \SI{30}{\gevc}$ in \pp{} collisions at $\sqrt{s} = \SI{13}{\tev}$, normalised to the total number of jets. Model/data ratios are shown in the bottom panels for PYTHIA 8~\cite{MONASH,PYTHIA6,PYTHIA8-2} and POWHEG~\cite{POWHEG} + PYTHIA~6~\cite{PYTHIA6} simulations.
    }
    \label{Fig:Final_Results_Zg}
\end{figure}

\begin{figure}[hpt]
    \centering
    \includegraphics[width=0.49\textwidth]{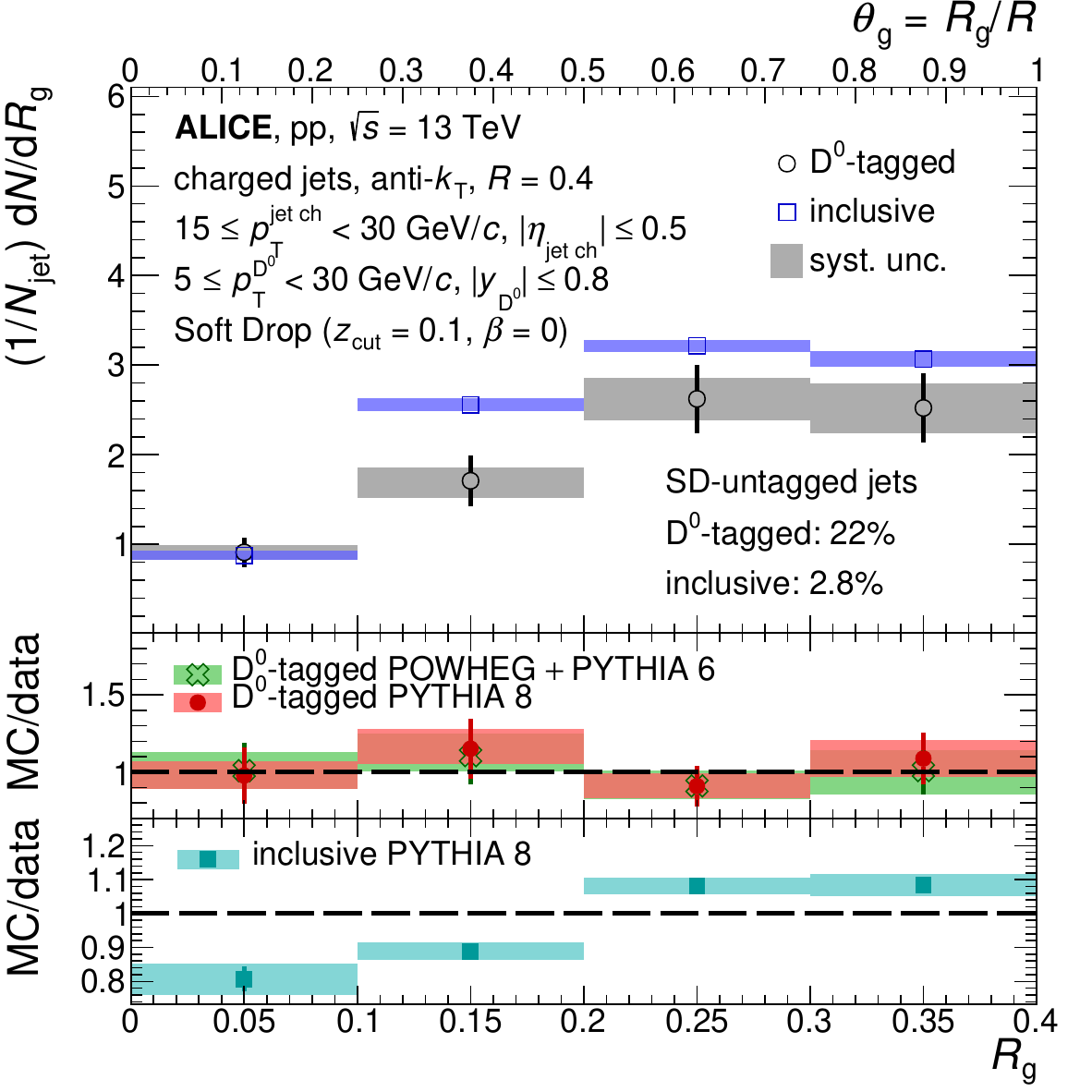}
    \includegraphics[width=0.49\textwidth]{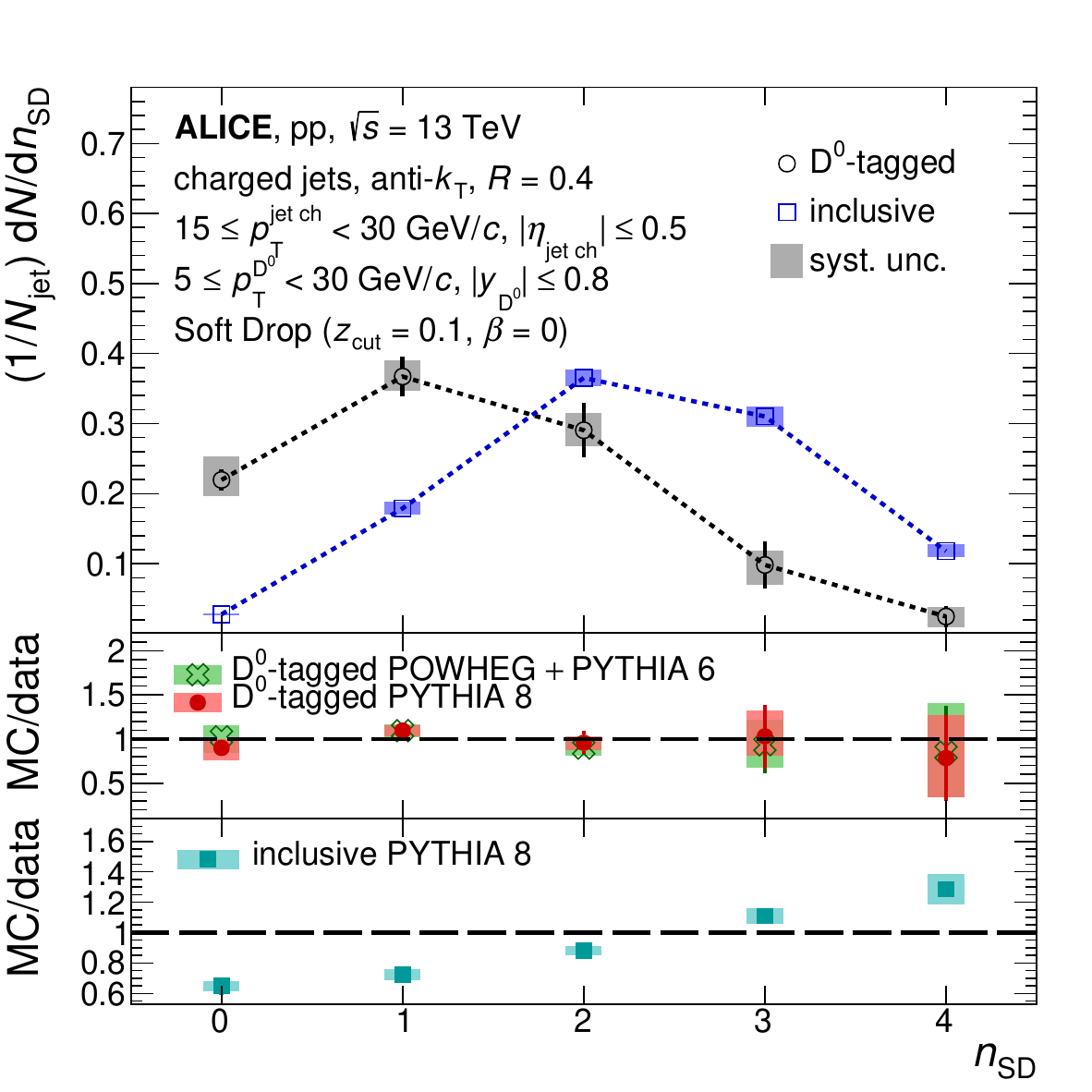}
    \caption{The $\rg$ (left) and $\nsd$ (right) distributions of prompt \dzero\hyp{}tagged jets compared to those of inclusive jets for $15 \leq \ptjet < \SI{30}{\gevc}$ in \pp{} collisions at $\sqrt{s} = \SI{13}{\tev}$. Model/data ratios are shown in the bottom panels for PYTHIA 8~\cite{MONASH,PYTHIA6,PYTHIA8-2} and POWHEG~\cite{POWHEG} + PYTHIA~6~\cite{PYTHIA6} simulations.
    }
    \label{Fig:Final_Results_Rg_Nsd}
\end{figure}

The first measurement mapping onto the charm-quark splitting function is reported via the $\zg$ distribution in Fig.~\ref{Fig:Final_Results_Zg}, for charm jets tagged by a~prompt \dzero{} meson. The $\rg$ and $\nsd$ distributions are shown in the left and right panels of Fig.~\ref{Fig:Final_Results_Rg_Nsd}, respectively. These distributions are fully corrected for detector effects and are reported in the jet transverse\hyp{}momentum interval of $15 \leq \ptjet < \SI{30}{\gevc}$.
Each charm-tagged jet measurement is accompanied by a measurement of an inclusive\hyp{}jet sample in the same kinematic interval, also fully corrected for detector effects. Both the charm-tagged jet and inclusive\hyp{}jet distributions are normalised to the total number of jets in each respective category, irrespective of whether they had a splitting passing the Soft Drop condition or not. In this way, the distributions are sensitive to the proportion of jets that do not have any splittings passing the Soft Drop condition, denoted as the SD\hyp{}untagged fraction. These fractions are \SI{22}{\percent} and \SI{2.8}{\percent}, in the charm-tagged jet and inclusive jet samples, respectively. Both samples are compared to PYTHIA~8 simulations and also to POWHEG + PYTHIA~6 simulations for the charm\hyp{}tagged jets. 
The ratios of predicted distributions from parton\hyp{}shower models to the measured distributions are also shown in the bottom panels of Figs.~\ref{Fig:Final_Results_Zg} and ~\ref{Fig:Final_Results_Rg_Nsd}.
A comparison of the measured $\zg$ distribution to analytical calculations performed in the soft-collinear effective theory (SCET) framework~\cite{vitev1,vitev2} can be found in Appendix~\ref{app:suppl}.

The $\zg$ distributions show that charm-tagged jets have significantly fewer symmetric splittings (large $\zg$ values), compared with inclusive jets. This is consistent with theoretical predictions~\cite{HFsplitting2017} of the role of mass effects in the QCD splitting function. 
The $\rg$ distribution for charm quarks shows a~reduction at large\hyp{}angles compared with inclusive jets. This can be due to the differences between quark and gluon fragmentation, with the gluon\hyp{}dominated inclusive sample expected to feature larger\hyp{}angle perturbative emissions. The charm and inclusive\hyp{}jet distributions are consistent at small $\rg$. This could be due to an interplay between the dead-cone effect, which suppresses small\hyp{}angle emissions from the charm quark, and the more abundant emissions from quarks compared with gluons at small angles, as gluons lose more of their initial virtuality through larger-angle emissions. 
The $\nsd$ distribution shows a~significant shift to smaller values for the charm\hyp{}tagged jets ($\langle\nsd\rangle = 1.34\pm0.08\text{ (stat.)}\pm0.07\text{ (syst.)}$) compared with inclusive jets ($\langle\nsd\rangle = 2.31\pm0.01\text{ (stat.)}\pm0.02\text{ (syst.)}$).
This indicates that, compared with light and massless partons, charm quarks on average emit fewer gluons with a large enough $\pt$ to pass the Soft Drop condition in the showering process. This is consistent with the expectation from the presence of a~dead cone for charm quarks, which results in a harder fragmentation of charm quarks compared with light quarks and gluons.

For all three observables, both the PYTHIA~8 and POWHEG + PYTHIA~6 predictions for charm-quark jets describe the measurements within uncertainties. 
For the more precise inclusive measurements, PYTHIA~8 exhibits some tension, particularly for the $\nsd$ observable. This may be partially due to the poorly constrained quark and gluon fractions in the inclusive sample at low $\ptjet$. In contrast, as the \dzero-tagged jets represent a quark-enriched sample, the comparison of the shower generators to data is less reliant on this fraction and can instead be used to more accurately study the mechanisms underpinning the shower properties.

The measurement of the groomed shared momentum fraction, $\zg$, of charm-tagged jets, reported in this Letter, represents the first direct experimental constraint of the splitting function of heavy-flavour quarks. The $\zg$ distribution appears steeper than that of light quarks and gluons, with a suppressed probability of symmetric emissions. Measurements of the number of splittings passing the Soft Drop condition as well as the emission angle of the first of these splittings, show that heavy-flavour quarks on average have fewer perturbative emissions compared with light quarks and gluons, with a lower probability of these emissions occurring at large angles. These different characteristics for heavy-quark emissions, compared with light quarks and gluons, constrain the roles of quark mass and Casimir colour factors in the parton shower, which are different between the two samples.
The upgraded ALICE detector in the LHC Run~3 will extend this measurement to jets tagged with a~fully reconstructed beauty meson~\cite{ALICE-PUBLIC-2020-005}. Direct comparison of the two quark-enriched samples of charm\hyp{}tagged and beauty\hyp{}tagged jets will enable the isolation of mass effects from the effects due to Casimir colour factors. The larger projected integrated luminosity in Run 3 will also allow us to extend the charm-tagged to inclusive jet comparisons to higher $\ptjet$, where mass effects become negligible and the remaining impact of the Casimir colour factors can be cleanly studied.


\newenvironment{acknowledgement}{\relax}{\relax}
\begin{acknowledgement}
\section*{Acknowledgements}

The ALICE Collaboration would like to thank all its engineers and technicians for their invaluable contributions to the construction of the experiment and the CERN accelerator teams for the outstanding performance of the LHC complex.
The ALICE Collaboration gratefully acknowledges the resources and support provided by all Grid centres and the Worldwide LHC Computing Grid (WLCG) collaboration.
The ALICE Collaboration acknowledges the following funding agencies for their support in building and running the ALICE detector:
A. I. Alikhanyan National Science Laboratory (Yerevan Physics Institute) Foundation (ANSL), State Committee of Science and World Federation of Scientists (WFS), Armenia;
Austrian Academy of Sciences, Austrian Science Fund (FWF): [M 2467-N36] and Nationalstiftung f\"{u}r Forschung, Technologie und Entwicklung, Austria;
Ministry of Communications and High Technologies, National Nuclear Research Center, Azerbaijan;
Conselho Nacional de Desenvolvimento Cient\'{\i}fico e Tecnol\'{o}gico (CNPq), Financiadora de Estudos e Projetos (Finep), Funda\c{c}\~{a}o de Amparo \`{a} Pesquisa do Estado de S\~{a}o Paulo (FAPESP) and Universidade Federal do Rio Grande do Sul (UFRGS), Brazil;
Bulgarian Ministry of Education and Science, within the National Roadmap for Research Infrastructures 2020-2027 (object CERN), Bulgaria;
Ministry of Education of China (MOEC) , Ministry of Science \& Technology of China (MSTC) and National Natural Science Foundation of China (NSFC), China;
Ministry of Science and Education and Croatian Science Foundation, Croatia;
Centro de Aplicaciones Tecnol\'{o}gicas y Desarrollo Nuclear (CEADEN), Cubaenerg\'{\i}a, Cuba;
Ministry of Education, Youth and Sports of the Czech Republic, Czech Republic;
The Danish Council for Independent Research | Natural Sciences, the VILLUM FONDEN and Danish National Research Foundation (DNRF), Denmark;
Helsinki Institute of Physics (HIP), Finland;
Commissariat \`{a} l'Energie Atomique (CEA) and Institut National de Physique Nucl\'{e}aire et de Physique des Particules (IN2P3) and Centre National de la Recherche Scientifique (CNRS), France;
Bundesministerium f\"{u}r Bildung und Forschung (BMBF) and GSI Helmholtzzentrum f\"{u}r Schwerionenforschung GmbH, Germany;
General Secretariat for Research and Technology, Ministry of Education, Research and Religions, Greece;
National Research, Development and Innovation Office, Hungary;
Department of Atomic Energy Government of India (DAE), Department of Science and Technology, Government of India (DST), University Grants Commission, Government of India (UGC) and Council of Scientific and Industrial Research (CSIR), India;
National Research and Innovation Agency - BRIN, Indonesia;
Istituto Nazionale di Fisica Nucleare (INFN), Italy;
Japanese Ministry of Education, Culture, Sports, Science and Technology (MEXT) and Japan Society for the Promotion of Science (JSPS) KAKENHI, Japan;
Consejo Nacional de Ciencia (CONACYT) y Tecnolog\'{i}a, through Fondo de Cooperaci\'{o}n Internacional en Ciencia y Tecnolog\'{i}a (FONCICYT) and Direcci\'{o}n General de Asuntos del Personal Academico (DGAPA), Mexico;
Nederlandse Organisatie voor Wetenschappelijk Onderzoek (NWO), Netherlands;
The Research Council of Norway, Norway;
Commission on Science and Technology for Sustainable Development in the South (COMSATS), Pakistan;
Pontificia Universidad Cat\'{o}lica del Per\'{u}, Peru;
Ministry of Education and Science, National Science Centre and WUT ID-UB, Poland;
Korea Institute of Science and Technology Information and National Research Foundation of Korea (NRF), Republic of Korea;
Ministry of Education and Scientific Research, Institute of Atomic Physics, Ministry of Research and Innovation and Institute of Atomic Physics and University Politehnica of Bucharest, Romania;
Ministry of Education, Science, Research and Sport of the Slovak Republic, Slovakia;
National Research Foundation of South Africa, South Africa;
Swedish Research Council (VR) and Knut \& Alice Wallenberg Foundation (KAW), Sweden;
European Organization for Nuclear Research, Switzerland;
Suranaree University of Technology (SUT), National Science and Technology Development Agency (NSTDA), Thailand Science Research and Innovation (TSRI) and National Science, Research and Innovation Fund (NSRF), Thailand;
Turkish Energy, Nuclear and Mineral Research Agency (TENMAK), Turkey;
National Academy of  Sciences of Ukraine, Ukraine;
Science and Technology Facilities Council (STFC), United Kingdom;
National Science Foundation of the United States of America (NSF) and United States Department of Energy, Office of Nuclear Physics (DOE NP), United States of America.
In addition, individual groups or members have received support from:
Marie Sk\l{}odowska Curie, European Research Council, Strong 2020 - Horizon 2020 (grant nos. 950692, 824093, 896850), European Union;
Academy of Finland (Center of Excellence in Quark Matter) (grant nos. 346327, 346328), Finland;
Programa de Apoyos para la Superaci\'{o}n del Personal Acad\'{e}mico, UNAM, Mexico.

\end{acknowledgement}

\bibliographystyle{utphys}   
\bibliography{bibliography}

\newpage
\appendix

\section{Supplementary material}
\label{app:suppl}

\begin{figure}[hpt]
    \centering
    \includegraphics[width=0.49\textwidth]{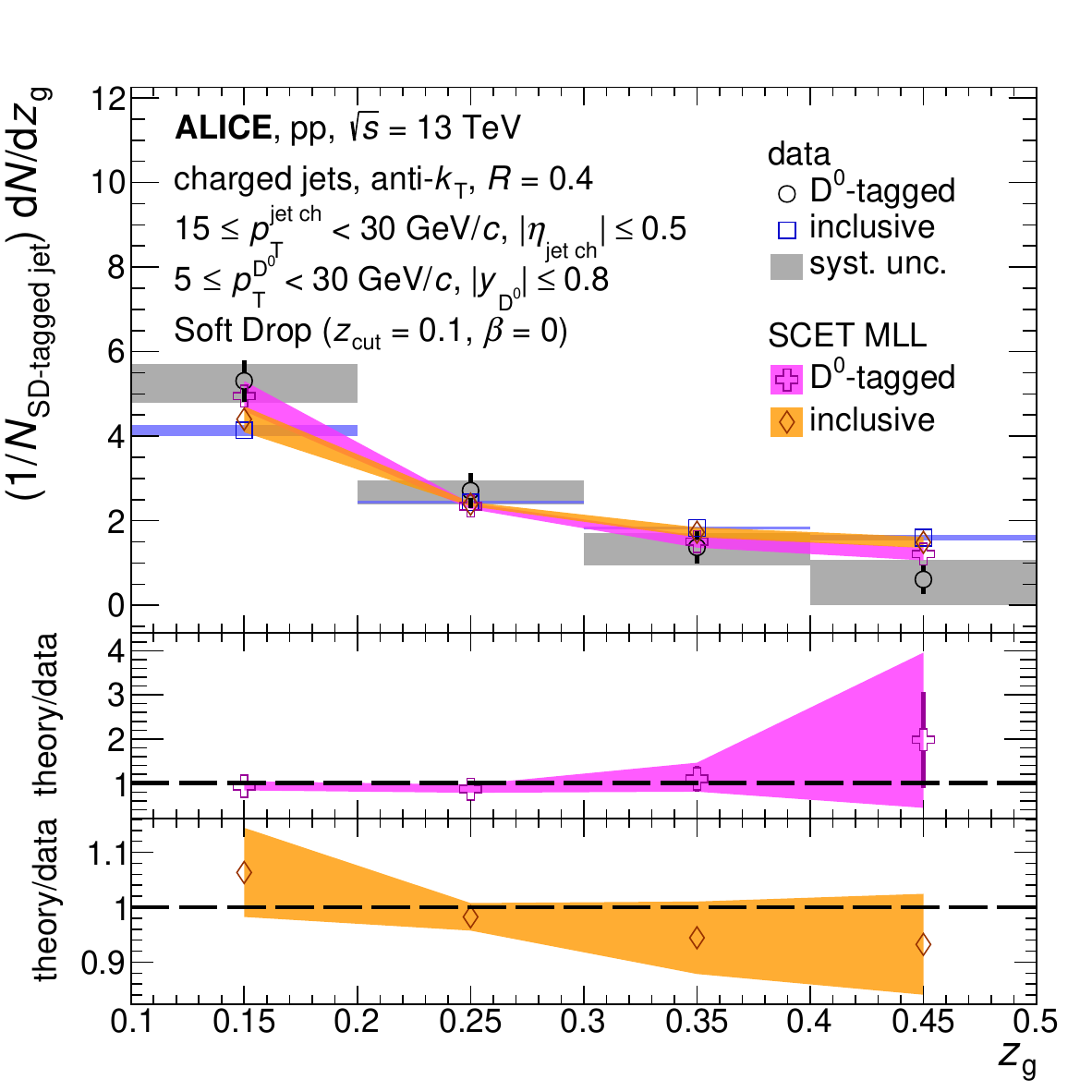}
    \caption{The $\zg$ distribution of prompt \dzero\hyp{}tagged jets compared to that of inclusive jets for \mbox{$15\leq\ptjet<\SI{30}{\gevc}$} in \pp{} collisions at $\sqrt{s} = \SI{13}{\tev}$, normalised to the number of jets passing the Soft Drop condition. Model/data ratios are shown in the bottom panels for soft-collinear effective theory calculations~\cite{vitev1,vitev2}.}
    \label{Fig:Final_Results_Zg_scet}
\end{figure}

The measured $\zg$ distribution is also compared to analytical calculations performed in the soft-collinear effective theory (SCET) framework with resummation to modified leading-logarithmic accuracy (MLL)~\cite{vitev1,vitev2}, presented in Fig.~\ref{Fig:Final_Results_Zg_scet}, which are consistent with the data. The calculations do not include hadronisation effects and are performed utilising a leading\hyp{}order evaluation of the charm cross section, which experimentally corresponds to jets with both charged and neutral information. It is expected that the low\hyp{}$\zg$ region of the calculations are affected by non-perturbative effects. For comparison with the SCET predictions, the measurement is normalised to the number of jets passing the Soft Drop condition in each sample, as the calculations are provided for $\zg \geq 0.1$ and do not include an untagged fraction.

\begin{figure}[hpt]
    \centering
    \includegraphics[width=0.49\textwidth]{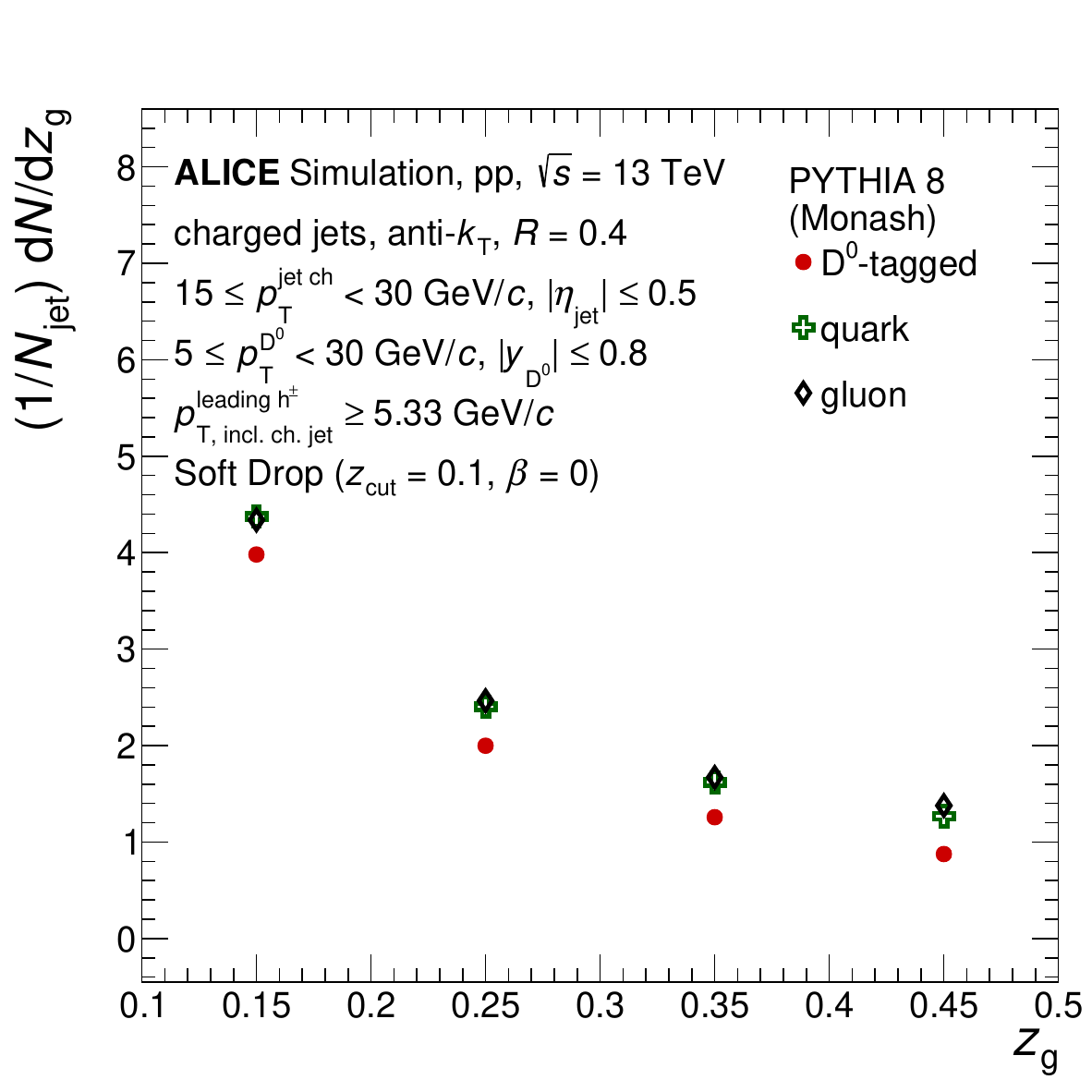}
    \caption{PYTHIA~8 $\zg$ distributions of prompt \dzero\hyp{}tagged jets, quark\hyp{}initiated and gluon\hyp{}initiated jets are compared, for \mbox{$15\leq\ptjet<\SI{30}{\gevc}$} in \pp{} collisions at $\sqrt{s} = \SI{13}{\tev}$.}
    \label{Fig:Zg_PYTHIA8}
\end{figure}

\begin{figure}[hpt]
    \centering
    \includegraphics[width=0.49\textwidth]{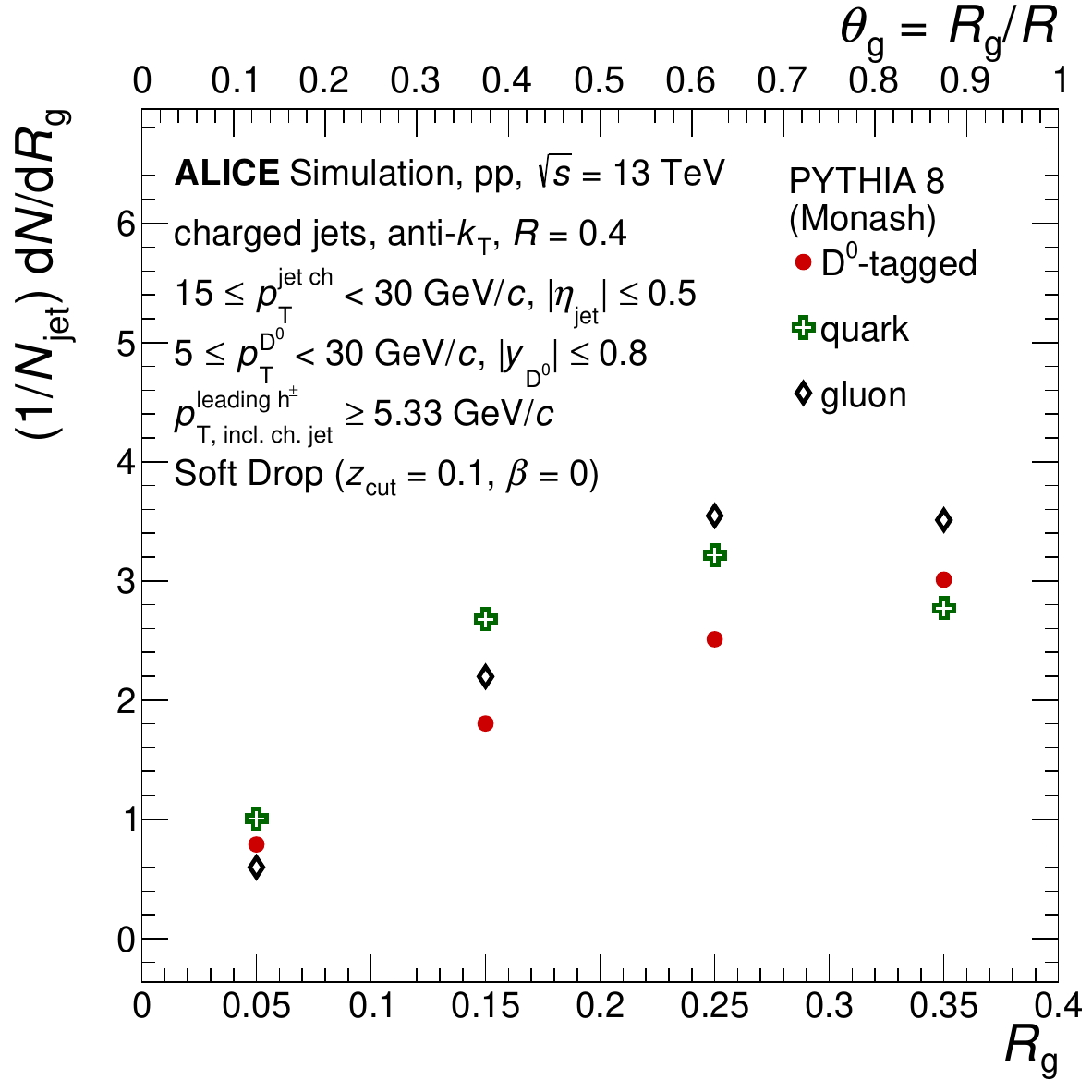}
    \caption{PYTHIA~8 $\rg$ distributions of prompt \dzero\hyp{}tagged jets, quark\hyp{}initiated and gluon\hyp{}initiated jets are compared, for \mbox{$15\leq\ptjet<\SI{30}{\gevc}$} in \pp{} collisions at $\sqrt{s} = \SI{13}{\tev}$.}
    \label{Fig:Rg_PYTHIA8}
\end{figure}

\begin{figure}[hpt]
    \centering
    \includegraphics[width=0.49\textwidth]{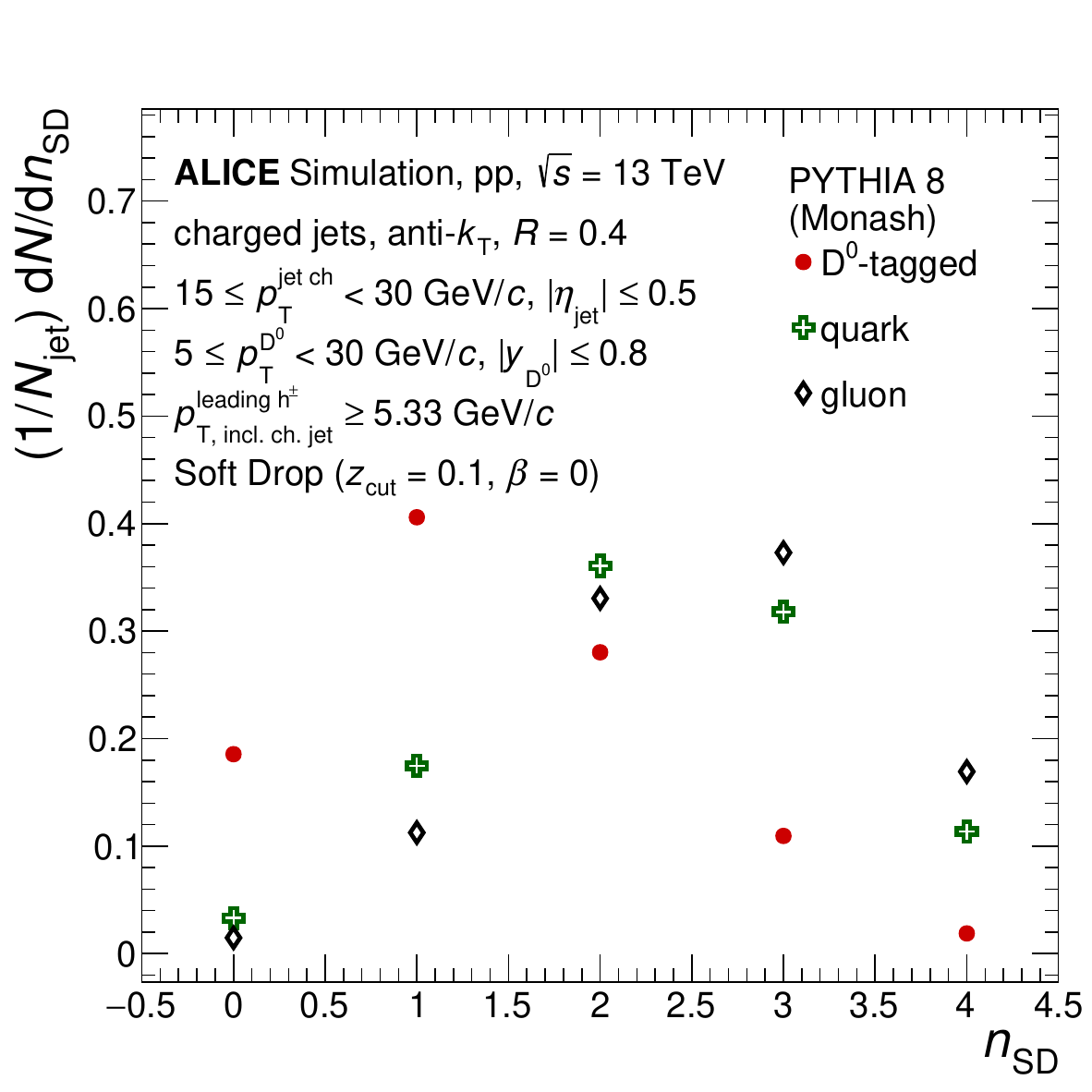}
    \caption{PYTHIA~8 $\nsd$ distributions of prompt \dzero\hyp{}tagged jets, quark\hyp{}initiated and gluon\hyp{}initiated jets are compared, for \mbox{$15\leq\ptjet<\SI{30}{\gevc}$} in \pp{} collisions at $\sqrt{s} = \SI{13}{\tev}$.}
    \label{Fig:Nsd_PYTHIA8}
\end{figure}

Comparisons of distributions of \dzero{}\hyp{}tagged jets, quark\hyp{}initiated jets and gluon\hyp{}initiated jets, obtained with PYTHIA~8 simulations, are shown in Figs.~\ref{Fig:Zg_PYTHIA8},~\ref{Fig:Rg_PYTHIA8} and~\ref{Fig:Nsd_PYTHIA8} for the $\zg$, $\rg$ and $\nsd$ observables, respectively. The distributions are all normalised to the total number of jets in the kinematic range, regardless of whether they had a splitting which successfully passed the Soft Drop condition. Quark (gluon)\hyp{}initiated jets are tagged by requiring an outgoing quark (gluon) from the initial hard scatter to be within $\Dt R < 0.4$ of the jet axis. Differences between the quark\hyp{}initiated jet distributions and the gluon\hyp{}initiated jet distributions highlight the role of Casimir colour effects in the observables, whilst differences between the quark\hyp{}initiated jet distributions and the \dzero{}\hyp{}tagged jet distributions are sensitive to mass effects. We observe that the $\zg$ observable is not very sensitive to the different Casimir colour factors of quarks and gluons, but shows a heightened sensitivity to mass effects. The $\rg$ distribution on the other hand is sensitive to both effects, with the larger Casimir colour factor of gluons compared to quarks resulting in broader opening angles for gluon\hyp{}initiated jets. The large mass of the charm quark also results in fewer small-angle emissions, compared to the quark\hyp{}initiated jet sample, as expected from the presence of a~large dead-cone region around the charm quark within which emissions are suppressed. The $\nsd$ distribution shows sensitivity to both Casimir and mass effects, with the impact of the latter being much more significant in this kinematic range. However, it is worth noting that flavour of emissions in the quark\hyp{}tagged and gluon\hyp{}tagged jets is only well defined for the first emission, whereas in the case of the \dzero{}\hyp{}tagged jets the flavour is controlled all the way through the shower.

\begin{figure}
    \centering
    \includegraphics[width=0.49\textwidth]{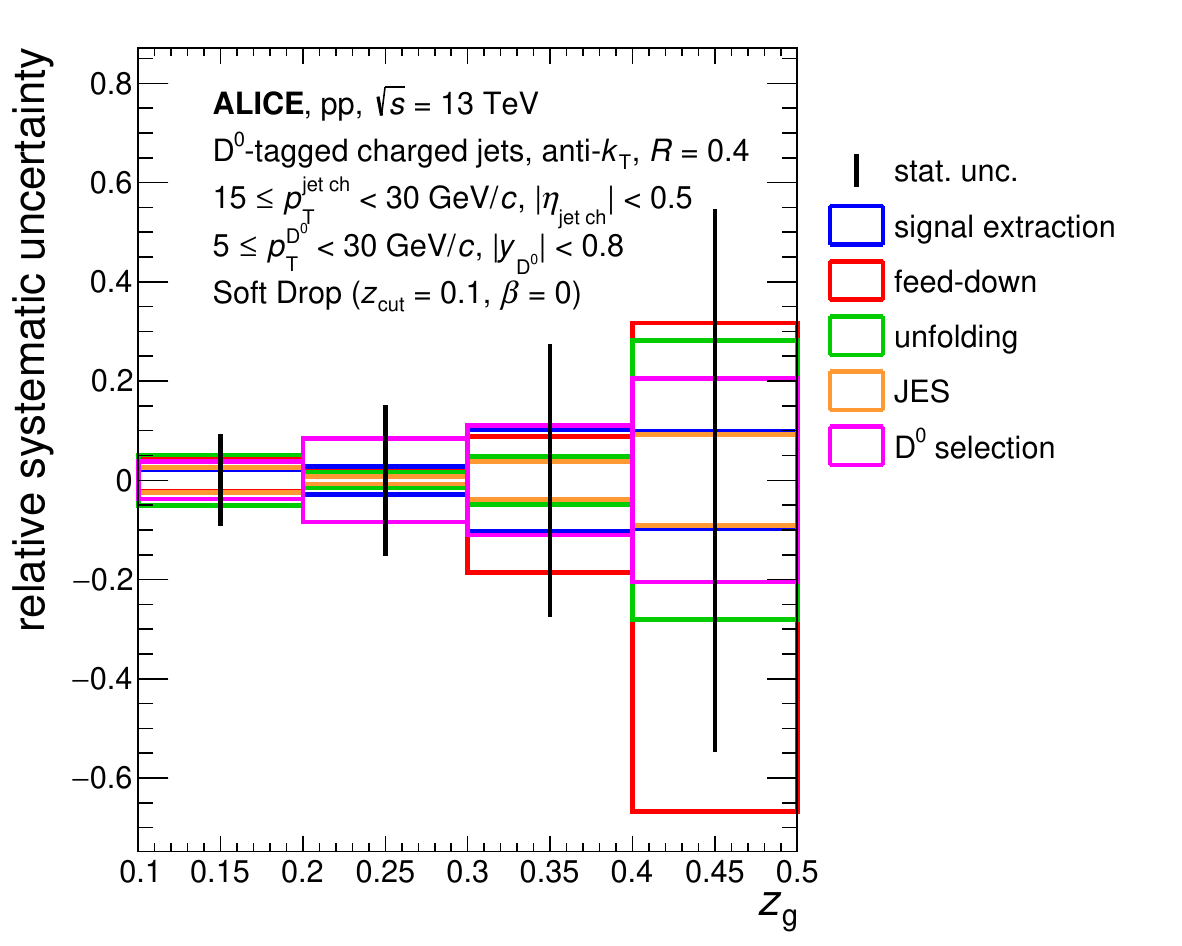}
    \caption{The systematic uncertainties of the measured $\zg$ distribution of prompt \dzero\hyp{}tagged jets for each uncertainty category, relative to the central value of the measurement in each interval, are presented. The statistical uncertainty relative to the central value is also shown.}
    \label{fig:syst_zg}
\end{figure}

\begin{figure}
    \centering
    \includegraphics[width=0.49\textwidth]{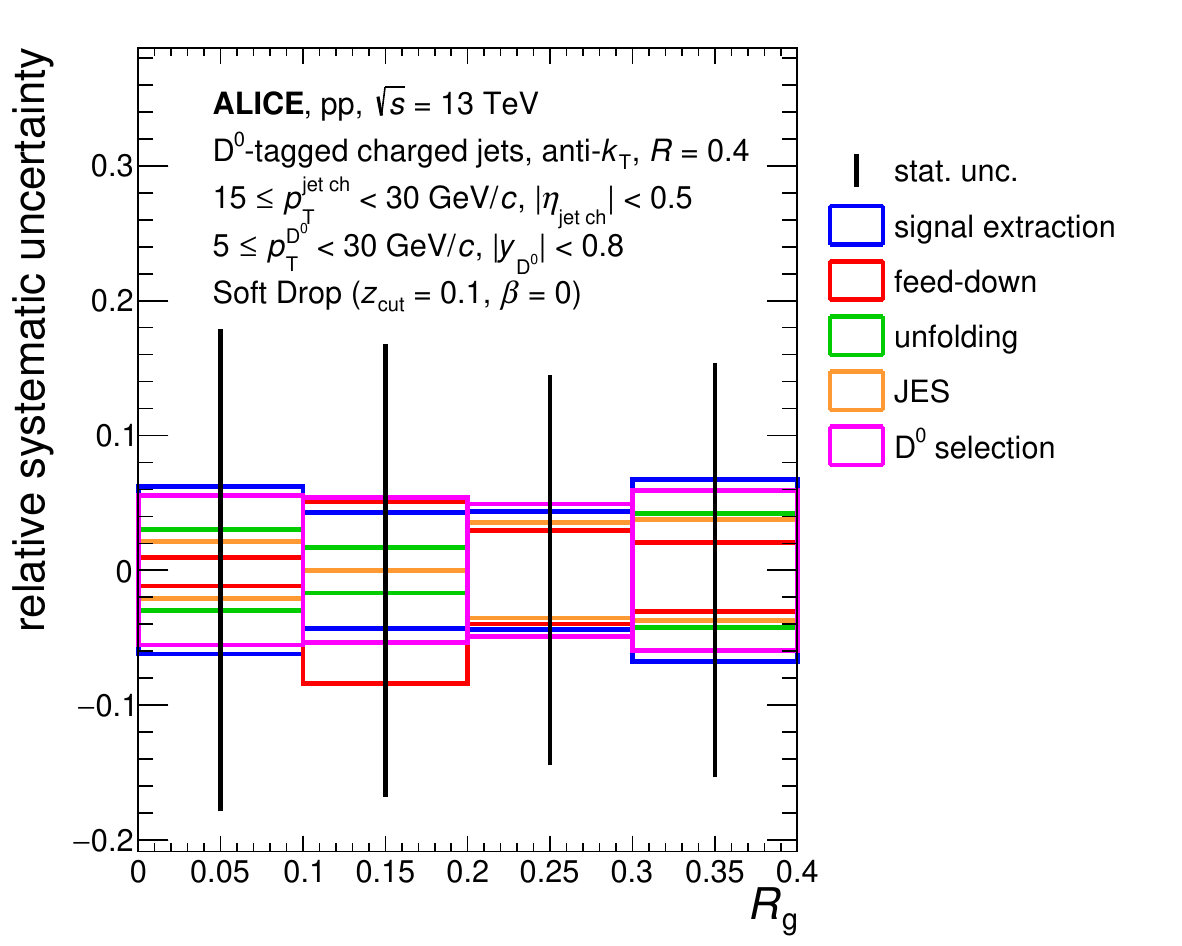}
    \includegraphics[width=0.49\textwidth]{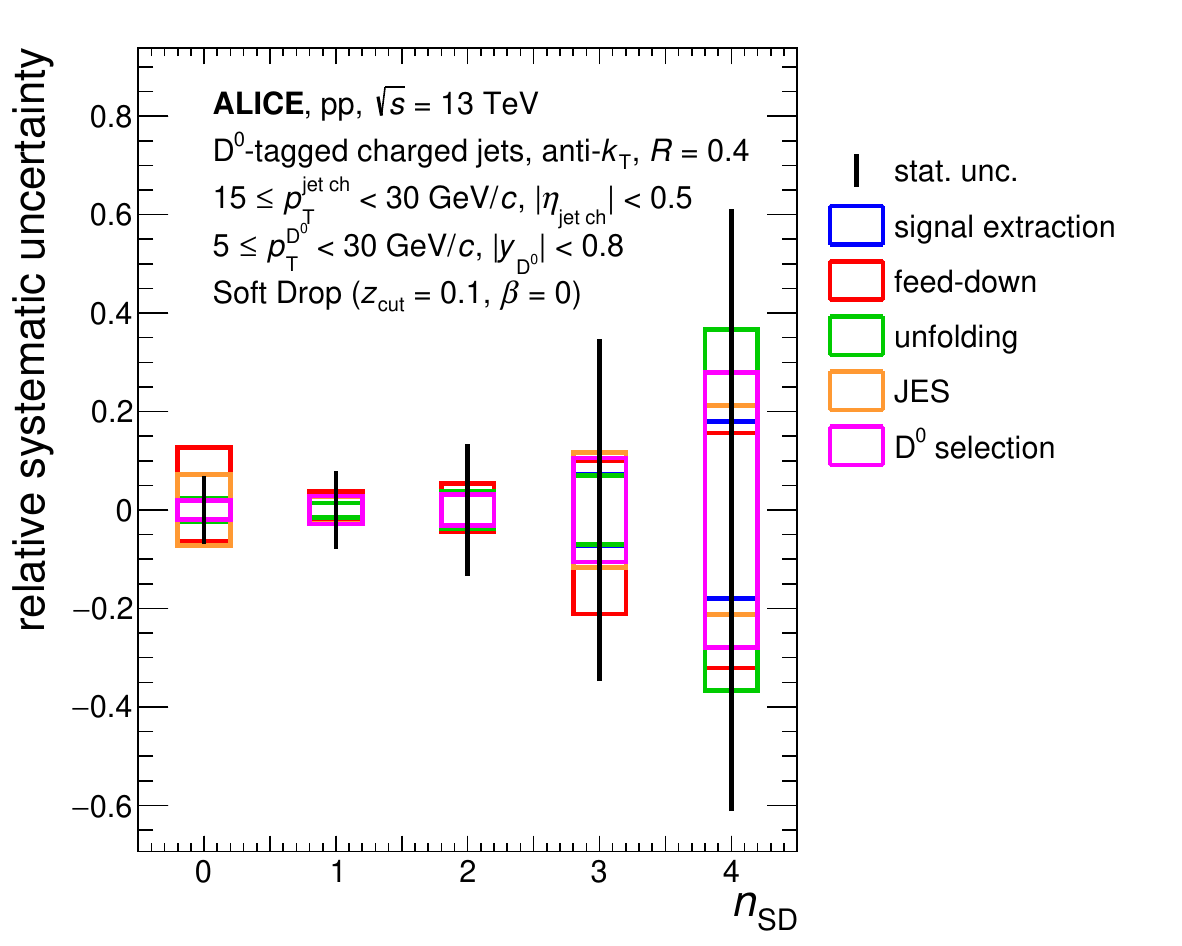}
    \caption{The systematic uncertainties of the measured $\rg$ (left) and $\nsd$ (right) distributions of prompt \dzero\hyp{}tagged jets for each uncertainty category, relative to the central value of the measurement in each interval, are presented. The statistical uncertainty relative to the central value is also shown.}
    \label{fig:syst_rg_nsd}
\end{figure}

The systematic uncertainties of the measured $\zg$, $\rg$ and $\nsd$ distributions of prompt \dzero\hyp{}tagged jets are presented in Figs.~\ref{fig:syst_zg} and~\ref{fig:syst_rg_nsd} for each individual uncertainty category considered (signal extraction, feed-down subtraction, unfolding, jet energy scale (JES), \dzero{} selection). The results are shown relative to the central values of the measured distributions in each interval. The statistical uncertainty relative to the central value is also shown for comparison.

\begin{figure}
    \centering
    \includegraphics[width=0.49\textwidth]{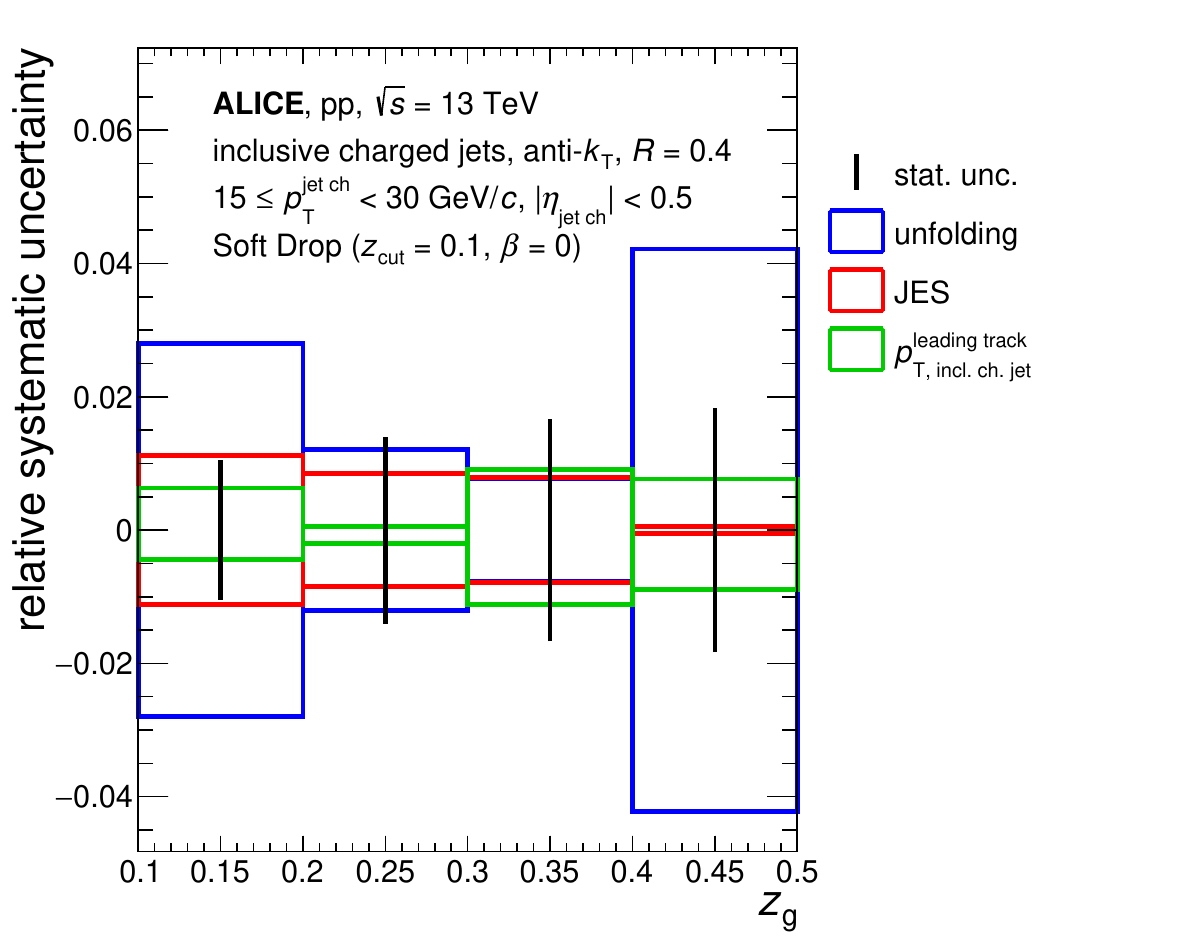}
    \caption{The systematic uncertainties of the measured $\zg$ distribution of inclusive jets for each uncertainty category, relative to the central value of the measurement in each interval, are presented. The statistical uncertainty relative to the central value is also shown.}
    \label{fig:syst_inclusive_zg}
\end{figure}

\begin{figure}
    \centering
    \includegraphics[width=0.49\textwidth]{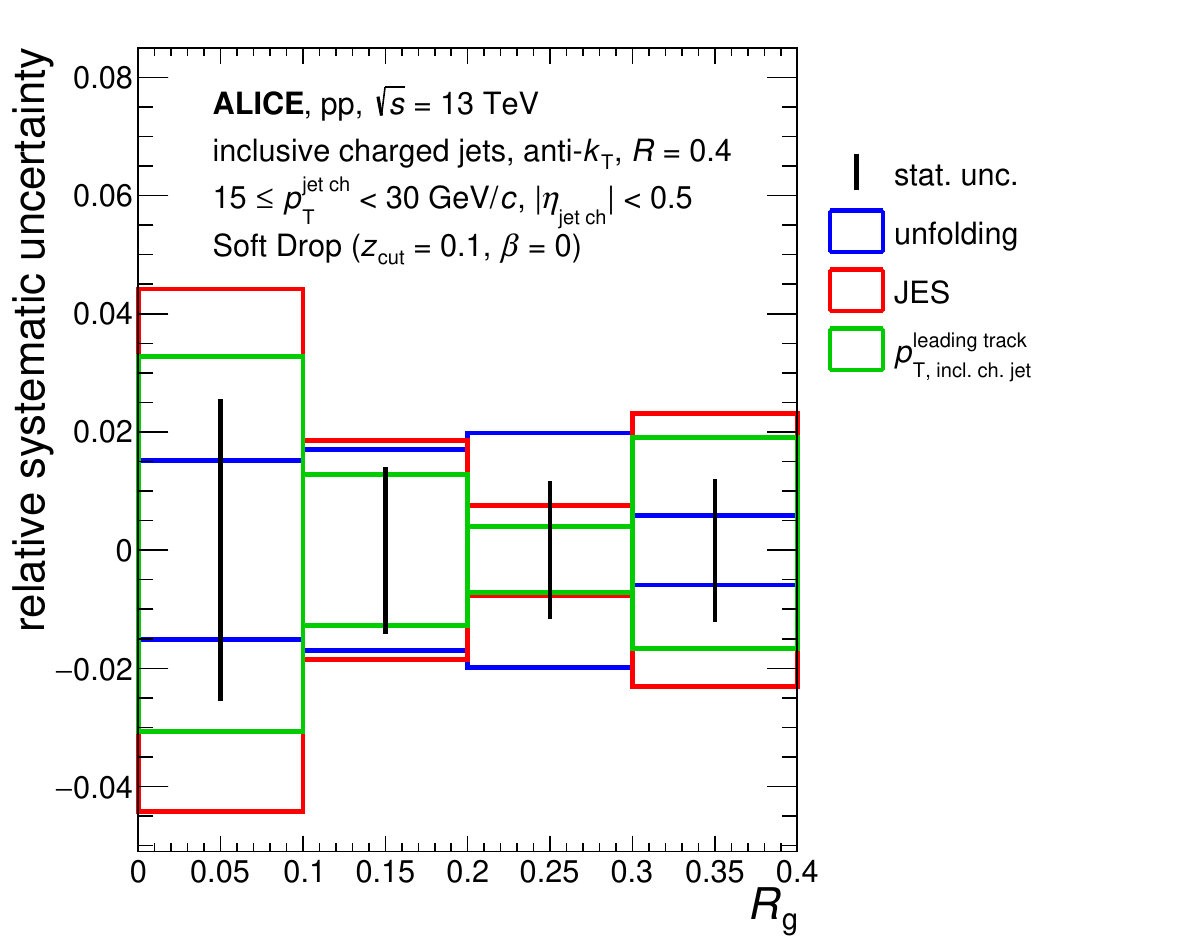}
    \includegraphics[width=0.49\textwidth]{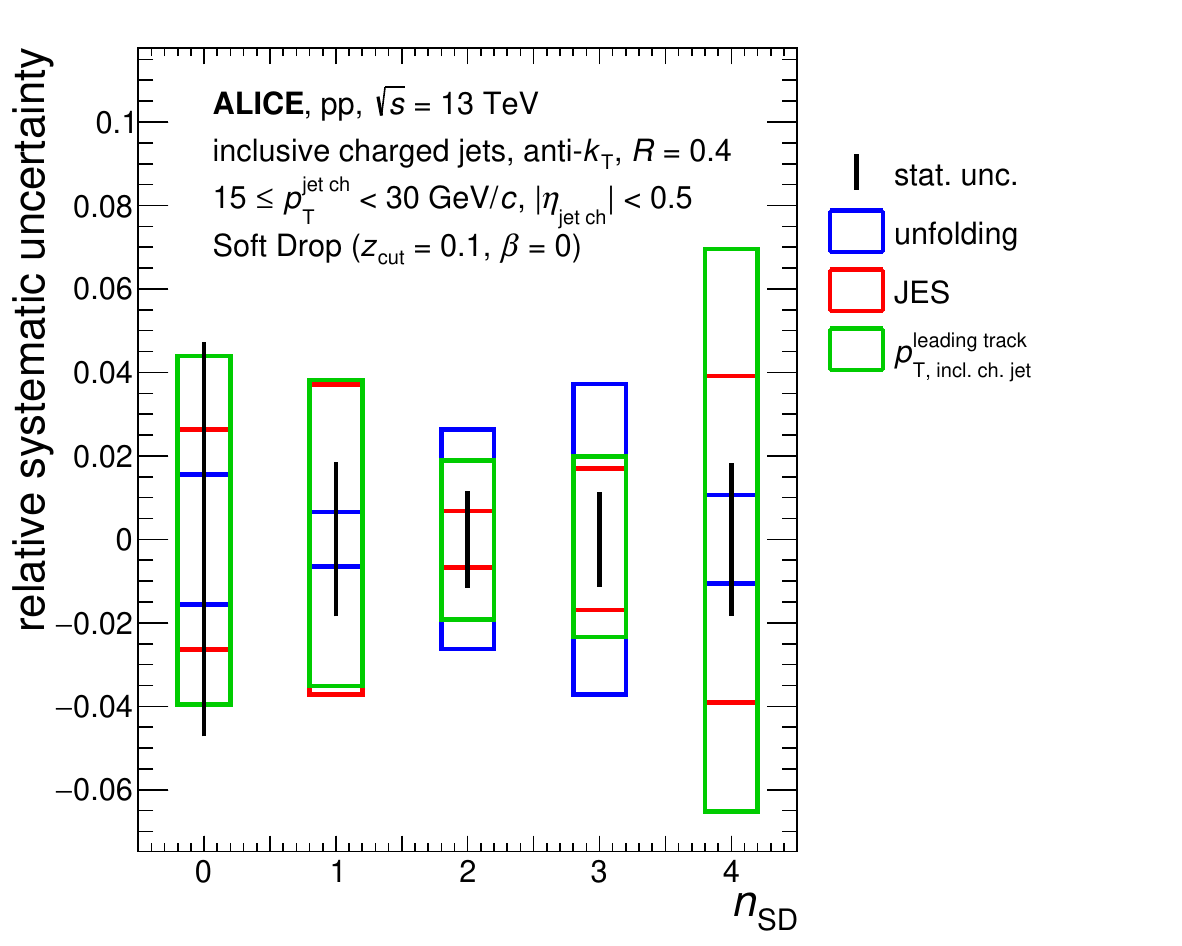}
    \caption{The systematic uncertainties of the measured $\rg$ (left) and $\nsd$ (right) distributions of inclusive jets for each uncertainty category, relative to the central value of the measurement in each interval, are presented. The statistical uncertainty relative to the central value is also shown.}
    \label{fig:syst_inclusive_rg_nsd}
\end{figure}

The systematic uncertainties of the measured $\zg$, $\rg$ and $\nsd$ distributions of inclusive jets are presented in Figs.~\ref{fig:syst_inclusive_zg} and~\ref{fig:syst_inclusive_rg_nsd} for each individual uncertainty category considered (unfolding, JES and $\pt$ selection on the leading track). The results are shown relative to the central values of the measured distributions in each interval. The statistical uncertainty relative to the central value is also shown for comparison.

\clearpage
%
%

\section{The ALICE Collaboration}
\label{app:collab}
\begin{flushleft} 
\small

S.~Acharya\,\orcidlink{0000-0002-9213-5329}\,$^{\rm 125}$, 
D.~Adamov\'{a}\,\orcidlink{0000-0002-0504-7428}\,$^{\rm 86}$, 
A.~Adler$^{\rm 69}$, 
G.~Aglieri Rinella\,\orcidlink{0000-0002-9611-3696}\,$^{\rm 32}$, 
M.~Agnello\,\orcidlink{0000-0002-0760-5075}\,$^{\rm 29}$, 
N.~Agrawal\,\orcidlink{0000-0003-0348-9836}\,$^{\rm 50}$, 
Z.~Ahammed\,\orcidlink{0000-0001-5241-7412}\,$^{\rm 132}$, 
S.~Ahmad\,\orcidlink{0000-0003-0497-5705}\,$^{\rm 15}$, 
S.U.~Ahn\,\orcidlink{0000-0001-8847-489X}\,$^{\rm 70}$, 
I.~Ahuja\,\orcidlink{0000-0002-4417-1392}\,$^{\rm 37}$
% ...

\section*{Affiliation Notes}

$^{\rm I}$ Deceased\\
$^{\rm II}$ Also at: Max-Planck-Institut f\"{u}r Physik, Munich, Germany\\
$^{\rm III}$ Also at: Italian National Agency for New Technologies, Energy and Sustainable Economic Development (ENEA), Bologna, Italy\\
$^{\rm IV}$ Also at: Dipartimento DET del Politecnico di Torino, Turin, Italy\\
$^{\rm V}$ Also at: Department of Applied Physics, Aligarh Muslim University, Aligarh, India\\
$^{\rm VI}$ Also at: Institute of Theoretical Physics, University of Wroclaw, Poland\\
$^{\rm VII}$ Also at: An institution covered by a cooperation agreement with CERN\\

\section*{Collaboration Institutes}

$^{1}$ A.I. Alikhanyan National Science Laboratory (Yerevan Physics Institute) Foundation, Yerevan, Armenia\\
$^{2}$ AGH University of Science and Technology, Cracow, Poland\\
$^{3}$ Bogolyubov Institute for Theoretical Physics, National Academy of Sciences of Ukraine, Kiev, Ukraine\\
$^{4}$ Bose Institute, Department of Physics  and Centre for Astroparticle Physics and Space Science (CAPSS), Kolkata, India\\
$^{5}$ California Polytechnic State University, San Luis Obispo, California, United States\\
$^{6}$ Central China Normal University, Wuhan, China\\
$^{7}$ Centro de Aplicaciones Tecnol\'{o}gicas y Desarrollo Nuclear (CEADEN), Havana, Cuba\\
$^{8}$ Centro de Investigaci\'{o}n y de Estudios Avanzados (CINVESTAV), Mexico City and M\'{e}rida, Mexico\\
$^{9}$ Chicago State University, Chicago, Illinois, United States\\
$^{10}$ China Institute of Atomic Energy, Beijing, China\\
$^{11}$ Chungbuk National University, Cheongju, Republic of Korea\\
$^{12}$ Comenius University Bratislava, Faculty of Mathematics, Physics and Informatics, Bratislava, Slovak Republic\\
$^{13}$ COMSATS University Islamabad, Islamabad, Pakistan\\
$^{14}$ Creighton University, Omaha, Nebraska, United States\\
$^{15}$ Department of Physics, Aligarh Muslim University, Aligarh, India\\
$^{16}$ Department of Physics, Pusan National University, Pusan, Republic of Korea\\
$^{17}$ Department of Physics, Sejong University, Seoul, Republic of Korea\\
$^{18}$ Department of Physics, University of California, Berkeley, California, United States\\
$^{19}$ Department of Physics, University of Oslo, Oslo, Norway\\
$^{20}$ Department of Physics and Technology, University of Bergen, Bergen, Norway\\
$^{21}$ Dipartimento di Fisica, Universit\`{a} di Pavia, Pavia, Italy\\
$^{22}$ Dipartimento di Fisica dell'Universit\`{a} and Sezione INFN, Cagliari, Italy\\
$^{23}$ Dipartimento di Fisica dell'Universit\`{a} and Sezione INFN, Trieste, Italy\\
$^{24}$ Dipartimento di Fisica dell'Universit\`{a} and Sezione INFN, Turin, Italy\\
$^{25}$ Dipartimento di Fisica e Astronomia dell'Universit\`{a} and Sezione INFN, Bologna, Italy\\
$^{26}$ Dipartimento di Fisica e Astronomia dell'Universit\`{a} and Sezione INFN, Catania, Italy\\
$^{27}$ Dipartimento di Fisica e Astronomia dell'Universit\`{a} and Sezione INFN, Padova, Italy\\
$^{28}$ Dipartimento di Fisica `E.R.~Caianiello' dell'Universit\`{a} and Gruppo Collegato INFN, Salerno, Italy\\
$^{29}$ Dipartimento DISAT del Politecnico and Sezione INFN, Turin, Italy\\
$^{30}$ Dipartimento di Scienze MIFT, Universit\`{a} di Messina, Messina, Italy\\
$^{31}$ Dipartimento Interateneo di Fisica `M.~Merlin' and Sezione INFN, Bari, Italy\\
$^{32}$ European Organization for Nuclear Research (CERN), Geneva, Switzerland\\
$^{33}$ Faculty of Electrical Engineering, Mechanical Engineering and Naval Architecture, University of Split, Split, Croatia\\
$^{34}$ Faculty of Engineering and Science, Western Norway University of Applied Sciences, Bergen, Norway\\
$^{35}$ Faculty of Nuclear Sciences and Physical Engineering, Czech Technical University in Prague, Prague, Czech Republic\\
$^{36}$ Faculty of Physics, Sofia University, Sofia, Bulgaria\\
$^{37}$ Faculty of Science, P.J.~\v{S}af\'{a}rik University, Ko\v{s}ice, Slovak Republic\\
$^{38}$ Frankfurt Institute for Advanced Studies, Johann Wolfgang Goethe-Universit\"{a}t Frankfurt, Frankfurt, Germany\\
$^{39}$ Fudan University, Shanghai, China\\
$^{40}$ Gangneung-Wonju National University, Gangneung, Republic of Korea\\
$^{41}$ Gauhati University, Department of Physics, Guwahati, India\\
$^{42}$ Helmholtz-Institut f\"{u}r Strahlen- und Kernphysik, Rheinische Friedrich-Wilhelms-Universit\"{a}t Bonn, Bonn, Germany\\
$^{43}$ Helsinki Institute of Physics (HIP), Helsinki, Finland\\
$^{44}$ High Energy Physics Group,  Universidad Aut\'{o}noma de Puebla, Puebla, Mexico\\
$^{45}$ Horia Hulubei National Institute of Physics and Nuclear Engineering, Bucharest, Romania\\
$^{46}$ Indian Institute of Technology Bombay (IIT), Mumbai, India\\
$^{47}$ Indian Institute of Technology Indore, Indore, India\\
$^{48}$ INFN, Laboratori Nazionali di Frascati, Frascati, Italy\\
$^{49}$ INFN, Sezione di Bari, Bari, Italy\\
$^{50}$ INFN, Sezione di Bologna, Bologna, Italy\\
$^{51}$ INFN, Sezione di Cagliari, Cagliari, Italy\\
$^{52}$ INFN, Sezione di Catania, Catania, Italy\\
$^{53}$ INFN, Sezione di Padova, Padova, Italy\\
$^{54}$ INFN, Sezione di Pavia, Pavia, Italy\\
$^{55}$ INFN, Sezione di Torino, Turin, Italy\\
$^{56}$ INFN, Sezione di Trieste, Trieste, Italy\\
$^{57}$ Inha University, Incheon, Republic of Korea\\
$^{58}$ Institute for Gravitational and Subatomic Physics (GRASP), Utrecht University/Nikhef, Utrecht, Netherlands\\
$^{59}$ Institute of Experimental Physics, Slovak Academy of Sciences, Ko\v{s}ice, Slovak Republic\\
$^{60}$ Institute of Physics, Homi Bhabha National Institute, Bhubaneswar, India\\
$^{61}$ Institute of Physics of the Czech Academy of Sciences, Prague, Czech Republic\\
$^{62}$ Institute of Space Science (ISS), Bucharest, Romania\\
$^{63}$ Institut f\"{u}r Kernphysik, Johann Wolfgang Goethe-Universit\"{a}t Frankfurt, Frankfurt, Germany\\
$^{64}$ Instituto de Ciencias Nucleares, Universidad Nacional Aut\'{o}noma de M\'{e}xico, Mexico City, Mexico\\
$^{65}$ Instituto de F\'{i}sica, Universidade Federal do Rio Grande do Sul (UFRGS), Porto Alegre, Brazil\\
$^{66}$ Instituto de F\'{\i}sica, Universidad Nacional Aut\'{o}noma de M\'{e}xico, Mexico City, Mexico\\
$^{67}$ iThemba LABS, National Research Foundation, Somerset West, South Africa\\
$^{68}$ Jeonbuk National University, Jeonju, Republic of Korea\\
$^{69}$ Johann-Wolfgang-Goethe Universit\"{a}t Frankfurt Institut f\"{u}r Informatik, Fachbereich Informatik und Mathematik, Frankfurt, Germany\\
$^{70}$ Korea Institute of Science and Technology Information, Daejeon, Republic of Korea\\
$^{71}$ KTO Karatay University, Konya, Turkey\\
$^{72}$ Laboratoire de Physique des 2 Infinis, Ir\`{e}ne Joliot-Curie, Orsay, France\\
$^{73}$ Laboratoire de Physique Subatomique et de Cosmologie, Universit\'{e} Grenoble-Alpes, CNRS-IN2P3, Grenoble, France\\
$^{74}$ Lawrence Berkeley National Laboratory, Berkeley, California, United States\\
$^{75}$ Lund University Department of Physics, Division of Particle Physics, Lund, Sweden\\
$^{76}$ Nagasaki Institute of Applied Science, Nagasaki, Japan\\
$^{77}$ Nara Women{'}s University (NWU), Nara, Japan\\
$^{78}$ National and Kapodistrian University of Athens, School of Science, Department of Physics , Athens, Greece\\
$^{79}$ National Centre for Nuclear Research, Warsaw, Poland\\
$^{80}$ National Institute of Science Education and Research, Homi Bhabha National Institute, Jatni, India\\
$^{81}$ National Nuclear Research Center, Baku, Azerbaijan\\
$^{82}$ National Research and Innovation Agency - BRIN, Jakarta, Indonesia\\
$^{83}$ Niels Bohr Institute, University of Copenhagen, Copenhagen, Denmark\\
$^{84}$ Nikhef, National institute for subatomic physics, Amsterdam, Netherlands\\
$^{85}$ Nuclear Physics Group, STFC Daresbury Laboratory, Daresbury, United Kingdom\\
$^{86}$ Nuclear Physics Institute of the Czech Academy of Sciences, Husinec-\v{R}e\v{z}, Czech Republic\\
$^{87}$ Oak Ridge National Laboratory, Oak Ridge, Tennessee, United States\\
$^{88}$ Ohio State University, Columbus, Ohio, United States\\
$^{89}$ Physics department, Faculty of science, University of Zagreb, Zagreb, Croatia\\
$^{90}$ Physics Department, Panjab University, Chandigarh, India\\
$^{91}$ Physics Department, University of Jammu, Jammu, India\\
$^{92}$ Physics Department, University of Rajasthan, Jaipur, India\\
$^{93}$ Physics Program and International Institute for Sustainability with Knotted Chiral Meta Matter (SKCM2), Hiroshima University, Hiroshima, Japan\\
$^{94}$ Physikalisches Institut, Eberhard-Karls-Universit\"{a}t T\"{u}bingen, T\"{u}bingen, Germany\\
$^{95}$ Physikalisches Institut, Ruprecht-Karls-Universit\"{a}t Heidelberg, Heidelberg, Germany\\
$^{96}$ Physik Department, Technische Universit\"{a}t M\"{u}nchen, Munich, Germany\\
$^{97}$ Politecnico di Bari and Sezione INFN, Bari, Italy\\
$^{98}$ Research Division and ExtreMe Matter Institute EMMI, GSI Helmholtzzentrum f\"ur Schwerionenforschung GmbH, Darmstadt, Germany\\
$^{99}$ Saga University, Saga, Japan\\
$^{100}$ Saha Institute of Nuclear Physics, Homi Bhabha National Institute, Kolkata, India\\
$^{101}$ School of Physics and Astronomy, University of Birmingham, Birmingham, United Kingdom\\
$^{102}$ Secci\'{o}n F\'{\i}sica, Departamento de Ciencias, Pontificia Universidad Cat\'{o}lica del Per\'{u}, Lima, Peru\\
$^{103}$ Stefan Meyer Institut f\"{u}r Subatomare Physik (SMI), Vienna, Austria\\
$^{104}$ SUBATECH, IMT Atlantique, Nantes Universit\'{e}, CNRS-IN2P3, Nantes, France\\
$^{105}$ Suranaree University of Technology, Nakhon Ratchasima, Thailand\\
$^{106}$ Technical University of Ko\v{s}ice, Ko\v{s}ice, Slovak Republic\\
$^{107}$ The Henryk Niewodniczanski Institute of Nuclear Physics, Polish Academy of Sciences, Cracow, Poland\\
$^{108}$ The University of Texas at Austin, Austin, Texas, United States\\
$^{109}$ Universidad Aut\'{o}noma de Sinaloa, Culiac\'{a}n, Mexico\\
$^{110}$ Universidade de S\~{a}o Paulo (USP), S\~{a}o Paulo, Brazil\\
$^{111}$ Universidade Estadual de Campinas (UNICAMP), Campinas, Brazil\\
$^{112}$ Universidade Federal do ABC, Santo Andre, Brazil\\
$^{113}$ University of Cape Town, Cape Town, South Africa\\
$^{114}$ University of Houston, Houston, Texas, United States\\
$^{115}$ University of Jyv\"{a}skyl\"{a}, Jyv\"{a}skyl\"{a}, Finland\\
$^{116}$ University of Kansas, Lawrence, Kansas, United States\\
$^{117}$ University of Liverpool, Liverpool, United Kingdom\\
$^{118}$ University of Science and Technology of China, Hefei, China\\
$^{119}$ University of South-Eastern Norway, Kongsberg, Norway\\
$^{120}$ University of Tennessee, Knoxville, Tennessee, United States\\
$^{121}$ University of the Witwatersrand, Johannesburg, South Africa\\
$^{122}$ University of Tokyo, Tokyo, Japan\\
$^{123}$ University of Tsukuba, Tsukuba, Japan\\
$^{124}$ University Politehnica of Bucharest, Bucharest, Romania\\
$^{125}$ Universit\'{e} Clermont Auvergne, CNRS/IN2P3, LPC, Clermont-Ferrand, France\\
$^{126}$ Universit\'{e} de Lyon, CNRS/IN2P3, Institut de Physique des 2 Infinis de Lyon, Lyon, France\\
$^{127}$ Universit\'{e} de Strasbourg, CNRS, IPHC UMR 7178, F-67000 Strasbourg, France, Strasbourg, France\\
$^{128}$ Universit\'{e} Paris-Saclay Centre d'Etudes de Saclay (CEA), IRFU, D\'{e}partment de Physique Nucl\'{e}aire (DPhN), Saclay, France\\
$^{129}$ Universit\`{a} degli Studi di Foggia, Foggia, Italy\\
$^{130}$ Universit\`{a} del Piemonte Orientale, Vercelli, Italy\\
$^{131}$ Universit\`{a} di Brescia, Brescia, Italy\\
$^{132}$ Variable Energy Cyclotron Centre, Homi Bhabha National Institute, Kolkata, India\\
$^{133}$ Warsaw University of Technology, Warsaw, Poland\\
$^{134}$ Wayne State University, Detroit, Michigan, United States\\
$^{135}$ Westf\"{a}lische Wilhelms-Universit\"{a}t M\"{u}nster, Institut f\"{u}r Kernphysik, M\"{u}nster, Germany\\
$^{136}$ Wigner Research Centre for Physics, Budapest, Hungary\\
$^{137}$ Yale University, New Haven, Connecticut, United States\\
$^{138}$ Yonsei University, Seoul, Republic of Korea\\
$^{139}$  Zentrum  f\"{u}r Technologie und Transfer (ZTT), Worms, Germany\\
$^{140}$ Affiliated with an institute covered by a cooperation agreement with CERN\\
$^{141}$ Affiliated with an international laboratory covered by a cooperation agreement with CERN.\\

\end{flushleft} 

\end{document}